\documentclass[preprint]{ptephy_v1}

\preprintnumber{YITP-16-56, MPP-2016-87, OU-HET-895} 

\usepackage{graphics} 

\newcommand{\mre}{\mathrm{e}}
\newcommand{\mrd}{\mathrm{d}}

\begin{document}

\title{Flow equation for the large $N$ scalar model and induced geometries}


\author{Sinya Aoki}
\affil{Center for Gravitational Physics, Yukawa Institute for Theoretical Physics, Kyoto University, 
Kitashirakawa Oiwakechou, Sakyo-ku, Kyoto 606-8502, Japan, and
Center for Computational Sciences, University of Tsukuba 305-8577, Japan
 \email{saoki@yukawa.kyoto-u.ac.jp}}

\author{Janos Balog}
\affil{Institute for Particle and Nuclear Physics, 
Wigner Research Centre for Physics, 
MTA Lend\"ulet Holographic QFT Group,
1525 Budapest 114, P.O.B.\ 49, Hungary \email{balog.janos@wigner.mta.hu}}

\author{Tetsuya Onogi} 
\affil{Department of Physics, Osaka University, Toyonaka, Osaka 560-0043, Japan \email{onogi@phys.sci.osaka-u.ac.jp}}

\author{Peter Weisz}
\affil{Max-Planck-Institut f\"ur Physik, 80805 Munich, Germany \email{pew@mpp.mpg.de}}


\begin{abstract}
\begin{center}
{\it We dedicate this work to the memory of Peter Hasenfratz. }\\
\end{center}
We study the proposal that a $d+1$ dimensional induced metric 
is constructed from a $d$ dimensional field theory using gradient flow. 
Applying the idea to the O($N$) $\varphi^4$ model and normalizing the flow 
field, we have shown in the large $N$ limit that the induced metric is finite 
and universal in the sense that it does not depend on the details of 
the flow equation and the original field theory except for the 
renormalized mass, which is the only relevant quantity in this limit. 
We have found that the induced metric describes Euclidean 
Anti-de-Sitter (AdS) space in both ultra-violet (UV) and infra-red (IR) 
limits of the flow direction, where the radius of the AdS is bigger 
in the IR than in the UV.
\end{abstract}

\subjectindex{B30, B32,B35,B37}

\maketitle

\section{Introduction}
\label{intro}
The AdS/CFT (or more generally Gravity/Gauge theory) 
correspondence\cite{Maldacena:1997re}
is a surprising but significant finding in field theories and string theories.
The original correspondence claims that a $d$ dimensional conformal field 
theory is equivalent to some $d+1$ dimensional super-gravity theory on 
an AdS background. 
After the first proposal, there appeared many pieces of
evidence which indicate that the correspondence is true,
and applications to various areas in physics, such as AdS/QCD 
or AdS/CMP, have been successfully investigated.
Even though the AdS/CFT correspondence might be explained by the closed 
string/open string duality, the claimed equivalence is still mysterious 
and a deeper understanding on the correspondence is necessary.   

In a previous paper\cite{Aoki:2015dla}, two of the present authors 
and their collaborator tried to understand the gravity/field theory 
correspondences from a different point of view,  proposing an 
alternative method to define a  geometry from a field theory:
A quantum field theory in $d$ dimensions is lifted to a $d+1$ 
dimensional one using gradient flow\cite{Narayanan:2006rf,Luscher:2010iy, Luscher:2009eq,Luscher:2013vga}, 
where the flow time $t$, which represents the energy scale of the original 
$d$ dimensional theory, is interpreted as an additional coordinate.  
Then the induced metric is defined from this $d+1$ dimensional field. 
As the metric is derived from the original $d$ dimensional theory 
together with its scale dependence, the method proposed in the paper 
can be applied in principle to all field theories.
As a concrete example, the method was applied to the O($N$) non-linear 
sigma model (NLSM) in two dimensions, and the vacuum expectation 
value (VEV) of the three dimensional induced metric was shown 
to describe an AdS space in the massless limit.\footnote{There are also studies using a different method on the relation between O(N) vector models in d-dim and (generalized) gravity theories in d+1 dim\cite{Klebanov:2002ja}. See also \cite{Giombi:2012ms} for recent developments.}

There are two key aspects in this proposal.  
First of all, the $d+1$ dimensional induced metric becomes classical 
in the large $N$ limit, and the quantum corrections can be calculated 
order by order in the large $N$ expansion.
Secondly, in the example mentioned above,
the three dimensional induced metric, constructed from a product of  
three dimensional flow fields at the same point, is free from  
UV divergences in the large $N$ limit. 
Instead, if the original two dimensional fields were used directly 
to define the two dimensional metric, it would badly diverge. 
These two special properties allow us to infer that (the VEV of) 
the induced metric describes a geometry.

In this paper, we further investigate this proposal, 
explicitly considering a more general model, the O($N$) invariant 
$\varphi^4$ model in $d$ dimensions, which can describe the free 
scalar model as well as the NLSM as limiting cases.  
Furthermore, as a generalization of the proposal in Ref.~\cite{Aoki:2015dla}, 
we consider the case that parameters (mass and coupling) are different 
in the original $d$ dimensional theory and the flow equation, 
in order to see how the behavior of the flow field depends on the choice.
We also introduce a normalization of the flow field, to interpret 
the gradient flow as a renormalization group (RG) transformation. 
Our normalization condition for the flow field requires a rescaling factor 
which becomes unity in the NLSM limit, so that the result in the 
previous paper\cite{Aoki:2015dla} is unchanged.
We then define the induced metric from this normalized flow field.
It turns out that the metric defined in this way leads to the same geometry, 
irrespective of parameters in the flow equation as 
well as in the original $d$ dimensional theory.
We show that the $d+1$ dimensional space described by the metric 
becomes AdS both in UV and IR limits as long as $d > 2$.

This paper is organized as follows. In Sec.~\ref{proposal}, 
we first briefly summarize the proposal in Ref.~\cite{Aoki:2015dla}, 
together with a few modifications introduced in this paper.
We explain some properties of the $d$ dimensional model in 
Sec.~\ref{model}, and solve the flow equation in the large $N$ limit in Sec.~\ref{flow}.
We calculate the induced metric in Sec.~\ref{geometry} and the Einstein 
tensor in Sec.~\ref{E-tensor}, showing that the space defined by 
the metric becomes AdS in both UV and IR limits. 
The behavior of the Einstein tensor between two limits is also 
investigated in Sec.~\ref{E-tensor}.
We finally give a summary of this paper in Sec.~\ref{summary}.
In appendix~\ref{app_divergence}, we show that a divergence of the flow field 
appears for the $\varphi^4$ theory in perturbation 
theory but it disappears non-perturbatively in the large $N$ limit. 
Some properties of the incomplete gamma function used in the main text  
are given in appendix~\ref{app_igamma}.

\section{Proposal}
\label{proposal}
In this section, we briefly summarize the proposal of 
Ref.~\cite{Aoki:2015dla} and explain some modifications introduced 
in this paper.

In this paper we consider the large $N$ scalar field $\varphi^{a}(x)$ 
with the $d$ dimensional space-time coordinate $x$ and the large $N$ 
index $a=1,2,\cdots, N $,  whose action $S$ describes the $\varphi^4$ 
model and will be explicitly given in the next section.

The $d$ dimensional field $\varphi^{ a}(x)$ is extended to $\phi^{ a}(t,x)$ 
in $d+1$ dimensions, using the gradient flow equation 
as\cite{Narayanan:2006rf,Luscher:2010iy} 
\begin{eqnarray}
\frac{\partial}{\partial t} \phi^{a}(t,x) &=& - \left. 
\frac{\delta S^\prime}{\delta \varphi^{a}(x)}
\right\vert_{\varphi \rightarrow\phi} , 
\label{eq:GFE}
\end{eqnarray}
with the initial condition $\phi^{a}(0,x)=\varphi^{a}(x)$, 
where the flow action $S^\prime$ in this paper can be different 
from the action $S$ in the original $d$ dimensional theory. 
Since the length dimension of $t$ is 2 and $t\ge 0$,  
a new variable $\tau =2\sqrt{t}$ is introduced.  
We then denote the $d+1$ dimensional coordinate as 
$z=(\tau,x)\, \in \mathbb {R}^+(=[0,\infty])\times \mathbb{R}^d$ 
and the flow field as $\phi^{a}(z)$.  

In Ref.~\cite{Aoki:2015dla}, a $d+1$ dimensional metric was defined as
\begin{eqnarray}
\hat g_{\mu\nu} (z) &\equiv &  h \sum_{a=1}^N \partial_\mu \phi^{a}(z) \partial_\nu \phi^{a}(z) ,
\label{eq:metric_old}
\end{eqnarray}
where $h>0$ is a dimensionful constant.
In this paper, however, we  modify this definition slightly  to
\begin{eqnarray}
\hat g_{\mu\nu} (z) &\equiv &  h \sum_{a=1}^N \partial_\mu \sigma^{a}(z) \partial_\nu \sigma^{a}(z) ,
\label{eq:metric}
\end{eqnarray}
where $\sigma^a$ is the normalized flow field defined by
\begin{eqnarray}
\sigma^a (z) &\equiv & \frac{\phi^a(z)}{\sqrt{\langle \phi^2(z)\rangle}} ,
\label{eq:rescaling}
\end{eqnarray} 
which is equivalent to  a constant renormalization, as we will see later,
and $h$ is a constant with the mass dimension $-2$. The average $\langle \phi^2(z) \rangle $ in the above definition will be defined in the next paragraph.
The reason why we use $\sigma^a$ instead of $\phi^a$ to define 
the metric is as follows.
The RG transformation consists of two procedures:  first UV modes are 
integrated out and then the field is normalized for the RG flow to have a 
fixed point. Since the gradient flow itself corresponds to the former 
procedure only, a field normalization is introduced as 
in eq.~(\ref{eq:rescaling}). One may adopt a different field normalization, 
but eq.~(\ref{eq:rescaling}) leads to interesting results as will be seen  
in later sections. We call this condition the NLSM normalization, 
as $\langle \sigma^2(z)\rangle = 1\,\,\forall z$.

The above $\hat g_{\mu\nu}(z)$ is an induced metric on a $d+1$ dimensional 
manifold $\mathbb{R}^+\times \mathbb{R}^d$ from some manifold in 
$\mathbb{R}^N$ defined by $\sigma^a(z)$, which becomes an 
$N-1$ dimensional sphere after the quantum average as 
$\langle \sigma^2(z)\rangle = 1$.
Here the expectation values of $\hat{g}_{\mu\nu}$ (and its correlations) 
are defined as
\begin{eqnarray}
\langle \hat g_{\mu\nu}(z) \rangle &\equiv&  \langle  \hat g_{\mu\nu}(z) \rangle_S , \\
\langle \hat g_{\mu_1\nu_1}(z_1)\cdots \hat g_{\mu_n\nu_n}(z_n) \rangle &\equiv& \langle \hat g_{\mu_1\nu_1}(z_1) \cdots \hat g_{\mu_n\nu_n}(z_n) \rangle_S, ~~~~
\end{eqnarray}
where $\langle {\cal O} \rangle_S$ is the expectation value of 
${\cal O}(\varphi)$ in $d$ dimensions with the action $S$ as
\begin{eqnarray}
\langle {\cal O} \rangle_S &\equiv& \frac{1}{Z}\int {\cal D}\varphi \, {\cal O}(\varphi)\, \mre^{-S}, 
\quad Z\equiv \int {\cal D}\varphi \,  \mre^{-S}\,. 
\end{eqnarray}

Even though the composite operator $\hat g_{\mu\nu}(z)$ contains 
a product of two local operators at the same point $z$, 
$\langle \hat g_{\mu\nu}(z)\rangle $ is finite as long as $\tau\not=0$ 
for the two dimensional NLSM in the large $N$ limit\cite{Aoki:2015dla}.
Although the finiteness of the flow field was proven for gauge 
theories\cite{Luscher:2011bx} and some scalar theories\cite{Makino:2014sta} 
(see also the latest extension in \cite{Hieda2016}),
it is not guaranteed in general. Indeed it was pointed out recently that 
the flow field of the $\varphi^4$ model gives an extra UV divergence  
in perturbation theory\cite{Capponi:2015ucc}.
(The modified flow equation can avoid this divergence.\cite{Fujikawa:2016qis})
As shown later, however, the induced metric $\hat g_{\mu\nu}$ in 
eq.~(\ref{eq:metric}) is free from such UV divergences in the large $N$ limit. 
In appendix~\ref{app_divergence}, we explicitly demonstrate that a 
divergence indeed appears in perturbation theory, but it disappears 
non-perturbatively in the large $N$ limit.  
On the other hand, if the $d$ dimensional induced metric were defined from 
the $d$ dimensional field $\varphi$, it would diverge badly, and hence a
geometry could not be defined from it.

Moreover, thanks to the large $N$ factorization, 
quantum fluctuations of the metric $\hat g_{\mu\nu}$ are suppressed 
in the large $N$ limit.
The $n$-point correlation function of $\hat g_{\mu\nu}$ behaves as
\begin{eqnarray}
\langle \hat g_{\mu_1\nu_1}(z_1)\cdots  \hat g_{\mu_n\nu_n}(z_n) \rangle &=&  
\prod_{i=1}^n
\langle \hat g_{\mu_i\nu_i}(z_i)  \rangle + O\left(\frac{1}{N}\right),
\end{eqnarray}
showing that the induced metric $\hat g_{\mu\nu}$ becomes classical in 
the large $N$ limit, and quantum fluctuations are sub-leading and calculable 
in the large $N$ expansion.
This property allows a geometrical interpretation of the metric $\hat g_{\mu\nu}$.
In Sec.~\ref{E-tensor}, the VEV of the Einstein tensor is calculated directly  
from  $\langle \hat g_{\mu\nu}\rangle$, as in the classical theory.
 
\section{Large $N$ model}
\label{model}
\subsection{Model}
In this paper, we consider  the  $N$ component scalar $\varphi^4$ model in $d$ dimensions, 
defined by the action
\begin{eqnarray}
S(\mu^2,u) &=&  N \int \mrd^d x\,\left[ \frac{1}{2}\partial^k \varphi (x)\cdot \partial_k \varphi (x) +  \frac{\mu^2}{2}\varphi^2(x) +\frac{u}{4!} \left(\varphi^2(x)\right)^2
\right] ,
\label{eq:action}
\end{eqnarray}
where $\varphi^a(x)$ is an $N$ component scalar field, $(\ \cdot \ )$ 
indicates an inner product of $N$ component vectors such that 
$\varphi^2(x) \equiv \varphi(x)\cdot\varphi(x) =\sum_{a=1}^N \varphi^a(x)\varphi^a(x)$, 
$\mu^2$  is the bare scalar mass parameter, and $u$ is the bare coupling 
constant of the $\varphi^4$ interaction, whose canonical dimension is $4-d$.

This model describes the free massive scalar at $u=0$, while 
it is equivalent to the NLSM in the $u\rightarrow \infty$ limit, 
whose action is obtained from eq.~(\ref{eq:action}) as
\begin{eqnarray}
S(\lambda) &=& \frac{ N}{2\lambda} \int \mrd^d x\,\partial^k \sigma (x)\cdot \partial_k \sigma (x) , \qquad
\sigma^2(x) = 1, 
\label{eq:action_NLSM}
\end{eqnarray}
with the replacement
\begin{eqnarray}
\sigma^a(x)&=&\sqrt{\lambda} \varphi^a(x), \qquad \lambda 
=  \lim_{u\rightarrow\infty}  -\frac{u}{6\mu^2} . 
\end{eqnarray}
In addition,  the flow time has to be also rescaled by $\lambda$ for the flow equation in this limit.

\subsection{Calculation in the large $N$ limit}

The partition function is evaluated as\cite{MoZi} 
\begin{eqnarray}
Z(J) &=& \int \left[{\cal D}\varphi\right] \mre^{-S(\mu^2,u)+(J, \varphi)} 
= \int  \left[{\cal D} \varphi\right] \left[{\cal D}\rho\right]
\delta\left(\rho-\varphi^2 \right) \mre^{-S(\mu^2,u)+(J, \varphi)} 
\nonumber \\
&=& \int  \left[{\cal D} \beta\right]
\exp\left[ - N S_{\rm eff}(\beta,J)\right],
\end{eqnarray}
where the effective action is given by
\begin{equation}
S_{\rm eff}(\beta, J)=\int \mrd^dx\,\left[ \frac{6}{u}\beta^2(x)
+ \frac{6\mu^2}{u}i\beta(x) \right]
+\frac{1}{2} {\rm Tr} \ln D[\beta] -\frac{1}{2N^2} ( J, D^{-1} J ) ,
\label{Seff}
\end{equation}
with $D[\beta] (x,y) \equiv \{-\nabla^2 + 2i\beta(x) \}\delta^{(d)}(x-y) $, and
\begin{eqnarray}
(F,G) \equiv  \int \mrd^dx\, F(x) G(x)
\end{eqnarray}
for arbitrary functions $F(x)$ and $G(x)$.

The large $N$ limit corresponds to the saddle point, $2 i\beta(x) = m^2$, determined  by the saddle point equation as
\begin{eqnarray}
\left.  \frac{\partial S_{\rm eff}(\beta, J)}{i\partial \beta(x)}\right\vert_{2i\beta(x)=m^2} &=& 
\frac{6}{u} (\mu^2 - m^2) + Z(m) = 0, 
 \label{eq:beta}
\end{eqnarray}
where
\begin{eqnarray}
Z(m) &=& \int \mrd p\, \frac{1}{p^2+m^2}\ge 0 , \qquad 
\mrd p \equiv\frac{\mrd^d p}{(2\pi)^d},
\end{eqnarray}
which is divergent at $d > 1$. 
We therefore introduce some regularization such as lattice, 
dimensional or momentum cut-off regularization, but do not specify 
it unless necessary. In this paper, sending the cut-off to infinity 
is called ``the continuum limit".

The two point function is then evaluated as 
\begin{eqnarray}
\langle \varphi^a(x) \varphi^b(y) \rangle &=& 
\delta^{ab}\frac{1}{N}\int \mrd p\, \frac{\mre^{ip(x-y)}}{p^2+m^2}
\end{eqnarray}
at $u\not=\infty$,
where  $m$ is regarded as the renormalized mass. Therefore $\mu^2$ must be 
tuned to keep $m^2$ finite in the continuum limit. 
The saddle point equation,
\begin{eqnarray}
\mu^2 &=& m^2 - \frac{u}{6}Z(m) ,
\end{eqnarray}
says that $\mu^2\rightarrow -\infty$ in the continuum limit at $d > 1$ 
as long as $u> 0$.

In the NLSM limit $u\rightarrow\infty$, we have
\begin{eqnarray}
\langle \sigma^a(x) \sigma^b(y) \rangle &=& 
\delta^{ab}\frac{\lambda}{N}\int \mrd p\, \frac{\mre^{ip(x-y)}}{p^2+m^2},
\end{eqnarray}
 where
 \begin{eqnarray}
 \frac{1}{\lambda} &=& Z(m) =  \int \mrd p\, \frac{1}{p^2+m^2}.
\end{eqnarray}
Therefore $\lambda \rightarrow 0$ in the continuum limit at $d>1$.

 \section{Flow equation and its solution in the large $N$ limit}
\label{flow}
\subsection{Flow equation}
In this paper, we consider the flow equation, given by
\begin{eqnarray}
\frac{\partial}{\partial t} \phi^a(t,x) &=& -\left. \frac{\delta S(\mu_f^2,u_f)}{\delta \varphi^a(x)}\right\vert_{\varphi\rightarrow\phi}
= \left(\Box -\mu_f^2\right)\phi^a(t,x) -\frac{u_f}{6}\phi^a(t,x)
\phi^2(t,x), 
\label{eq:flow} \\
\phi^a(0,x) &=&\varphi^a(x), \nonumber
\end{eqnarray}
where $\mu_f^2$ and $u_f$ can be different from $\mu^2$ and $u$ in 
the original $d$-dimensional theory.
The flow with $\mu_f=\mu$ and $u_f=u$ is called gradient flow, as it is 
given by the gradient of the original action. 

\subsection{The solution in the large $N$ limit}
In the large $N$ limit, the solution has the following form in momentum 
space\cite{Aoki:2014dxa},
\begin{eqnarray}
\phi(t,p) &=& f(t) \mre^{-p^2 t} \varphi(p).
\label{eq:sol_N}
\end{eqnarray}
The flow equation (\ref{eq:flow}) in the large $N$ limit leads 
to the equation for $f(t)$:
\begin{eqnarray}
\dot f(t) &=& -\mu_f^2 f(t) -\frac{u_f}{6} f^3(t) \zeta_0(t), 
\label{eq:eq_f}
\end{eqnarray}
where
\begin{eqnarray}
\zeta_0(t) &=& \int \mrd p\, \frac{\mre^{-2p^2 t}}{p^2+m^2} , \qquad \zeta_0(0) = Z(m).
\end{eqnarray}
Using $X(t) = f^{-2}(t)$, the equation becomes  
\begin{eqnarray}
\dot X(t) &=& 2\mu_f^2 X(t) +\frac{u_f}{3} \zeta_0(t), 
\label{eq:eq_X}
\end{eqnarray}
so that the solution is given by
\begin{eqnarray}
X(t) &=& \mre^{2t \mu_f^2}\left[1+\frac{u_f}{3} \int_0^t  \mrd x\, 
\zeta_0(x)\mre^{-2x\mu_f^2 }\right] .
\end{eqnarray}
Introducing $m_f$ by the relation
\begin{eqnarray}
\mu^2_f&=& m_f^2 -\frac{u_f}{6} Z(m_f) ,
\end{eqnarray}
the solution is further reduced as
\begin{eqnarray}
X(t)&=&\frac{\mre^{2t \mu_f^2}}{m_f^2-\mu_f^2}\left[m_f^2-\mu_f^2 + \frac{u_f}{3} \int_0^t  \mrd x\,\int \mrd p\,
\left(
\frac{p^2+m_f^2}{p^2+m^2}-\frac{p^2+\mu_f^2}{p^2+m^2}
\right)
 \mre^{-2x(p^2+\mu_f^2) }\right] \nonumber \\
 &=& \frac{\zeta(t)}{Z(m_f)}, \qquad \zeta(t)\equiv  \zeta_0(t) +\Delta(t), 
\label{eq:sol_X}
\end{eqnarray}
where
\begin{eqnarray}
\Delta (t) &=& \left\{ Z(m_f) - Z(m) \right\} \mre^{2t\mu_f^2} 
+ \int \mrd p\, \left(\frac{p^2+m_f^2}{p^2+m^2}\right)\frac{\mre^{2t\mu_f^2} 
- \mre^{-2t p^2}}{p^2+\mu_f^2} .
\label{eq:def_Delta}
\end{eqnarray}

In the case of the interacting flow with $u_f > 0$, 
$\mu_f^2$ negatively diverges ($\mu_f^2\rightarrow -\infty$),
as $Z(m_f)\rightarrow +\infty $ (the continuum limit  at $d > 1$)
or $u_f\rightarrow +\infty$ (the NLSM limit).  
Therefore $\Delta(t)$ vanishes as 
\begin{eqnarray}
\lim_{\mu_f^2\rightarrow -\infty} \Delta (t) &\simeq & -\frac{m_f^2\zeta_0(t)-\dot\zeta_0(t)/2}{\mu_f^2} + O\left(\mu_f^{-4}\right)
\label{eq:lim_Delta}
\end{eqnarray}
for $t > 0$.

For the flow with $u_f=0$, which we call ``the free flow", 
we simply have $f(t)=\mre^{-m_f^2t}$ for all $t\ge 0$.

\subsection{Two-point function}
The two-point function of the flow field with $u_f > 0$ is given by
\begin{eqnarray}
\langle \phi^a(t,x)\phi^b(s,y) \rangle &= & 
\frac{\delta^{ab}}{N}\frac{Z(m_f)}{\sqrt{\zeta(t)\zeta(s)}}
\int \mrd p\, \frac{\mre^{-(t+s)p^2}\mre^{ip(x-y)}}{p^2+m^2} .
\label{eq:2pt}
\end{eqnarray}
In the continuum or NLSM limit, we have
\begin{eqnarray}
\langle \phi^a(t,x)\phi^b(s,y) \rangle &= & 
\frac{\delta^{ab}}{N}\frac{Z(m_f)}{\sqrt{\zeta_0(t)\zeta_0(s)}}\int \mrd p\, 
\frac{\mre^{-(t+s)p^2}\mre^{ip(x-y)}}{p^2+m^2},
\label{eq:lim_2pt}
\end{eqnarray}
which only depends on renormalized quantities $m_f$ and $m$, 
but does not depend on bare parameters, $\mu_f^2$, $u_f$, $\mu^2$ and $u$. 
Note however that $Z(m_f)$ diverges in the continuum limit for $d > 1$. 

If we take the NLSM limit for the flow equation ($u_f\rightarrow\infty$) 
but without taking the continuum limit,
the two-point function of $\sigma^a(t,x) = \sqrt{\lambda_f} \phi^a(t,x)$ 
with $\lambda_f= 1/Z(m_f)$ is given by
\begin{eqnarray}
\langle \sigma^a(t,x)\sigma^b(s,y) \rangle &= & \frac{\delta^{ab}}{N}\frac{1}{\sqrt{\zeta_0(t)\zeta_0(s)}}\int \mrd p\, 
\frac{\mre^{-(t+s)p^2}\mre^{ip(x-y)}}{p^2+m^2}. 
\label{eq:2pt_NLSM}
\end{eqnarray}
This two-point function is finite even in the continuum limit, as shown 
explicitly at $d=2$\cite{Aoki:2014dxa,Makino:2014cxa}.
 
At $u_f=0$, on the other hand, we obtain
\begin{eqnarray}
\langle \phi^a(t,x)\phi^b(s,y) \rangle &= & \frac{\delta^{ab}}{N} 
\int \mrd p\, \frac{\mre^{-(t+s)(p^2+m_f^2)}\mre^{ip(x-y)}}{p^2+m^2} ,
\label{eq:2pt_free}
\end{eqnarray}
which is of course manifestly finite.

\subsection{Normalized flow field}
As mentioned before, to construct a RG transformation using 
the flow equation, which is merely a kind of smearing procedure for UV 
fluctuations, we introduce a normalization condition for the flow field. 

There is no unique way to define the normalized flow field. In the standard 
block spin transformation, for example, the normalization factor for the 
field is so chosen so that some fixed point can appear for the defined 
RG transformation.  
In this paper, we propose the following normalization for the flow field:
\begin{eqnarray}
 \sigma^a(t,x) &= & \frac{\phi^a(t,x)}{\sqrt{\langle \phi^2(t,x)\rangle} }, 
\label{eq:rescaled_flow}
\end{eqnarray}
so that the normalized flow field is dimensionless and satisfies the 
non-linear constraint $\langle \sigma^2(t,x)\rangle = 1$.
Because of this property, we may call it the NLSM normalization.

For the interacting flow ($u_f\not=0$), since
\begin{eqnarray}
\langle \phi^2(t,x)\rangle &=&\frac{Z(m_f)}{\zeta(t)} \int \mrd p\, 
\frac{\mre^{-2tp^2}}{p^2+m^2}  =Z(m_f) \frac{\zeta_0(t)}{\zeta(t)},
\label{eq:normalization}
\end{eqnarray}
we have
\begin{eqnarray}
\sigma^a(t,p) &=&\frac{1}{\sqrt{\zeta_0(t)}}\mre^{-p^2t} \varphi^a(p),
\label{eq:rescale_flow2}
\end{eqnarray}
which gives
\begin{eqnarray}
\langle \sigma^a(t,x)\sigma^b(s,y) \rangle &= & \frac{\delta^{ab}}{N}\frac{1}{\sqrt{\zeta_0(t)\zeta_0(s)}}\int \mrd p\, \frac{\mre^{-(t+s)p^2}\mre^{ip(x-y)}}{p^2+m^2}. 
\label{eq:2pt_rescale}
\end{eqnarray}
This result is not only finite in the continuum limit but also independent 
of bare parameters ($\mu_f^2, u_f,\mu^2,u$) as well as the renormalized 
flow mass $m_f$ even without taking the continuum or NLSM limit.

For the free flow ($u_f=0$), we have
\begin{eqnarray}
\langle \phi^2(t,x)\rangle &=& \int \mrd p\, \frac{\mre^{-2{ t}(p^2+m_f^2)}}{p^2+m^2}  = \zeta_0(t)\mre^{-2t m_f^2},
\label{eq:normalization_free}
\end{eqnarray}
which also leads to eq.~(\ref{eq:2pt_rescale}).

We here conclude that the normalized flow field defined by 
eq.~(\ref{eq:rescaled_flow}) leads universally to 
eqs.~(\ref{eq:rescale_flow2}) and (\ref{eq:2pt_rescale}), 
which do not depend on flow parameters as well as bare parameters 
$\mu^2$ and $u$ in the original theory. The final result is universal 
and depends only on the renormalized mass $m$ in the original theory. 
 Even  the free massless  flow equation ($u_f=m_f=0$)
gives the same result as long as the normalized flow field is used.
Note also that the flow field in the NLSM automatically leads 
to eqs.~(\ref{eq:rescale_flow2}) and (\ref{eq:2pt_rescale})
without introducing the non-trivial normalization factor.

\section{Induced metric and geometry}
\label{geometry}
\subsection{Induced Metric}
As proposed in Sec.~\ref{proposal}, we define the symmetric 2nd-rank tensor field as
\begin{eqnarray}
\hat g_{\mu\nu}(z) &\equiv& h \sum_{a=1}^N \partial_\mu\sigma^a(z) \partial_\nu \sigma^a (z), 
\end{eqnarray}
where  
$\mu,\nu$ run from 0 to $d$, and
$h$ is a constant with mass dimension $-2$.
We can interpret this field as the induced metric on the $d+1$ dimensional 
manifold $\mathbb{R}^+ \times \mathbb{R}^d$ from some manifold in 
$\mathbb{R}^d$ defined by $\sigma^a(z)$, which classically becomes
the $N-1$ dimensional sphere. 

Using eq.~(\ref{eq:2pt_rescale}), the VEV of $\hat g_{\mu\nu}$ is given by 
\begin{eqnarray}
g_{\mu\nu}(z) &\equiv& \langle \hat g_{\mu\nu}(z)\rangle = \left(
\begin{array}{cc}
 g_{\tau\tau}(\tau) & 0     \\
 0 & g_{ij} (\tau)     \\
\end{array}
\right)
\end{eqnarray}
where
\begin{eqnarray}
g_{\tau\tau}(\tau) = \frac{h \tau^2}{16} 
\frac{\mrd^2 \log \zeta_0(t)}{\mrd t^2}, \qquad
g_{ij}(\tau) &=& -\delta_{ij}\frac{h}{2d} \frac{\mrd \log \zeta_0(t)}{\mrd t}.
\label{eq:gmunu}
\end{eqnarray}
Explicitly $\zeta_0(t)$ is evaluated as
\begin{eqnarray}
\zeta_0(t) &=& \frac{m^{d-2} \mre^{2m^2t}}{(4\pi)^{d/2}} \Gamma(1-d/2, 2m^2 t) .
\end{eqnarray}
See appendix \ref{app_igamma} for the definition and properties of the incomplete gamma function $\Gamma(a,x)$.

\subsection{The metric in the IR limit}
In the IR limit $m\tau \gg 1$, we have
\begin{eqnarray}
\zeta_0(t) \simeq \frac{1}{(4\pi)^{d/2} m^2 (2t)^{d/2}},
\end{eqnarray}
so that
\begin{eqnarray}
g_{\tau\tau}(\tau) =\frac{h d}{2\tau^2} , \qquad
g_{ij}(\tau)&=& \frac{h\delta_{ij}}{\tau^2} .
\end{eqnarray}
This result shows that $g_{\mu\nu}$, the VEV of the induced metric 
$\hat g_{\mu\nu}$, describes a Euclidean AdS space in $d+1$ dimensions 
at $m\tau \gg 1$ for any $d$. Indeed, with a new variable 
$u = \sqrt{d/2}\tau$, the world line element is expressed as 
\begin{eqnarray}
\mrd s^2 &=& \frac{h d}{2 u^2}(\mrd u^2 + \mrd x^2) .
\end{eqnarray}
The condition that $\mrd s^2 >0$ with the Euclidean signature requires $h > 0$.

\subsection{The metric in the UV limit}
In the opposite limit $m\tau\ll 1$, on the other hand, we obtain,
\begin{eqnarray}
\zeta_0(t) &\simeq& \left\{
\begin{array}{lc}
  \displaystyle\frac{1}{2m} -\frac{(2t)^{1/2}}{\sqrt{\pi}}, &   d=1   \\
  \\
 \displaystyle  -\frac{\log (2t m^2) }{4\pi}, & d=2    \\
\\
\displaystyle \frac{2}{d-2}  \frac{(2t)^{1-d/2}}{(4\pi)^{d/2}}, &     d\ge 3
\end{array}
\right. ,
\end{eqnarray}
which leads to
\begin{eqnarray}
g_{\tau\tau}(\tau) &\simeq&
h \left\{
\begin{array}{lc}
  \displaystyle\sqrt{\frac{2}{\pi}}  \frac{m}{4\tau}, &   d=1   \\
  \\
 \displaystyle  -\frac{1 }{\tau^2\log (m^2\tau^2)}, & d=2    \\
\\
\displaystyle \frac{d-2}{2}  \frac{1}{\tau^2}.  &     d\ge 3
\end{array}
\right. ,
\end{eqnarray}
\begin{eqnarray}
g_{ij}(\tau) &\simeq& h\delta_{ij} \left\{
\begin{array}{lc}
  \displaystyle\sqrt{\frac{2}{\pi}}  \frac{m}{\tau}, &   d=1   \\
  \\
 \displaystyle  -\frac{1 }{\tau^2\log (m^2\tau^2)}, & d=2    \\
\\
\displaystyle \frac{d-2}{d}  \frac{1}{\tau^2}, &     d\ge 3
\end{array}
\right. .
\end{eqnarray}
The VEV of the metric describes a Euclidean AdS space at $d\ge 3$, 
while a log correction appears at $d=2$.

\section{Einstein Tensor}
\label{E-tensor}
In this section, we consider the VEV of  the Einstein tensor $G_{\mu\nu}$. 
As mentioned in Sec.~\ref{proposal}, quantum corrections 
can be neglected in the large $N$ limit as
\begin{eqnarray}
\langle G_{\mu\nu}(\hat g_{\mu\nu} ) \rangle  = G_{\mu\nu}(\langle \hat g_{\mu\nu} \rangle ) +
O\left(\frac{1}{N}\right) .
\end{eqnarray}
Therefore, the VEV of the Einstein tensor $G_{\mu\nu}$ can be calculated 
from the VEV of the induced metric, 
$g_{\mu\nu}=\langle \hat g_{\mu\nu}\rangle$, in this limit.

\subsection{Einstein tensor from $g_{\mu\nu}$}
If the metric has the following simple form  
\begin{eqnarray}
g_{\mu\nu}(\tau) &=&\left(
\begin{array}{ccc}
B(\tau)  & 0     \\
0  &   \delta_{ij} A(\tau)    \\
\end{array}
\right),
\label{eq:metric_diag}
\end{eqnarray}
then the Einstein tensor $G_{\mu\nu} = R_{\mu\nu} - R g_{\mu\nu}/2$ becomes
\begin{eqnarray}
G_{\mu\nu}(\tau) &=&\left(
\begin{array}{ccc}
G_{\tau\tau}  & 0     \\
0  &   G_{ij}    \\
\end{array}
\right),
\label{eq:Gmunu}
\end{eqnarray}
where
\begin{eqnarray}
G_{\tau\tau} &=&  \frac{d(d-1)}{8}\frac{{\dot A}^2}{A^2}, \qquad 
\dot f\equiv \frac{\mrd f(\tau)}{\mrd\tau}\,, \\
G_{ij} &=& \delta_{ij} \frac{(d-1)}{4}\left[\frac{(d-4)}{2}\frac{{\dot A}^2}{AB} + 2\frac{\ddot A}{B} - \frac{\dot A \dot B}{B^2}\right] .
\end{eqnarray}

From eq.~(\ref{eq:gmunu}), we have
\begin{eqnarray}
A(\tau) &=& - \frac{h}{d}  m^2 Y(x), \quad 
B(\tau) = \frac{h}{4}m^2 (m\tau)^2 \frac{\mrd Y(x)}{\mrd x}
\end{eqnarray}
with $x=m^2\tau^2/2$, where
\begin{eqnarray}
Y(x) &=& \frac{\mrd }{\mrd x}\left[\log\left\{ 
\frac{m^{d-2} \mre^{x}}{(4\pi)^{d/2}}\Gamma(1-d/2,x)\right\} \right]
= 1+\frac{\mrd }{\mrd x} \log \Gamma(1-d/2,x).
\end{eqnarray}
After a little algebra, we obtain 
\begin{eqnarray}
G_{\tau\tau}(\tau) &=& \frac{d(d-1)m^2}{8}(m\tau)^2 
\left[ \frac{\mrd \log Y(x)}{\mrd x} \right]^2_{x=m^2\tau^2/2}, \\
G_{ij}(\tau) &=& - \delta_{ij} \frac{(d-1)m^2}{d} \left[ \frac{(d-4)}{2}
\frac{\mrd \log Y(x)}{\mrd x} + \frac{\mrd}{\mrd x}
\left\{ \log\left(\frac{\mrd Y(x)}{\mrd x} \right)\right\} \right]_{x=m^2\tau^2/2} ,
\end{eqnarray}
so that $G_{\mu\nu}=0$ at $d=1$.

Two scalar functions $\Lambda_{\tau}(m\tau)$ and $\Lambda_{d}(m\tau)$,  
defined by 
\begin{eqnarray}
G_{\tau\tau}(\tau) =- \Lambda_\tau (m\tau) g_{\tau\tau}(\tau), \qquad
G_{ij}(\tau) &=& -\Lambda_d(m\tau) g_{ij}(\tau), 
\end{eqnarray}
become
\begin{eqnarray}
\Lambda_\tau(m\tau) &=&\left.  -\frac{d(d-1)}{2h} 
\frac{\displaystyle\frac{\mrd \log Y(x)}{\mrd x}}{Y(x)} \right\vert_{x=m^2\tau^2/2}, 
\label{eq:Lambda_d}\\
\Lambda_d(m\tau) &=& \Lambda_\tau( m\tau) + \delta\Lambda( m\tau) , \qquad
\frac{\delta \Lambda (m\tau)}{\Lambda_\tau(m\tau)} = \frac{2}{d}
\left[ \frac{\displaystyle \frac{\mrd}{\mrd x}\log
\left(\frac{\mrd Y(x)}{\mrd x}\right)}
{\displaystyle\frac{\mrd \log Y(x)}{\mrd x}}-2\right]_{x=m^2\tau^2/2} .~~
\label{eq:delta_Lambda}
\end{eqnarray}

\subsection{IR behaviors}
For $m\tau\gg 1$, we have
\begin{eqnarray}
\Lambda_\tau (m\tau) 
&\simeq& -\frac{(d-1)}{h}\left[1 + \frac{2(d+2)}{(m^2\tau^2)^2} +\cdots\right] ,~~~
\label{eq:Lambda_tau_IR} \\
\frac{\delta \Lambda (m\tau)}{\Lambda_\tau(m\tau)}
&\simeq& \frac{(2d+4)}{d}\frac{4}{ (m^2\tau^2)^2} .
\label{eq:del_Lambda_tau_IR}
\end{eqnarray}

Using the above results, we obtain
\begin{eqnarray}
G_{\mu\nu}(\tau) &\simeq& -\Lambda g_{\mu\nu} (\tau), \quad m\tau \gg 1, 
\end{eqnarray}
where  the cosmological constant is given by $\Lambda= -(d-1)/h$. 
Therefore the space defined by the metric $g_{\mu\nu}$ is the Euclidean AdS,  
except for $d=1$ where the space is flat ($\Lambda=0$).
This is consistent with the result in the previous section.

\subsection{UV limit}
We next consider the behavior of $G_{\mu\nu}$ in the  UV limit ($m \tau \ll 1$).

At $d=2$,  we have
\begin{eqnarray}
\Lambda_\tau(m\tau)&\simeq&\frac{1}{h}\left(L+1-2TL^2 -T\right), \qquad
\frac{\delta \Lambda (m\tau)}{\Lambda_\tau(m\tau)} \simeq  -\frac{L}{(L+1)^2}, \qquad m\tau \ll 1, 
\end{eqnarray}
where $T\equiv m^2\tau^2/2$ and $L=\log(T) +\gamma$.
Therefore we obtain
\begin{eqnarray}
G_{\mu\nu}(\tau) &\simeq & -\Lambda(m\tau) g_{\mu\nu}(\tau), \qquad m\tau \ll 1,
\end{eqnarray}
where the scalar function $\Lambda(\tau)$ has a logarithmic singularity 
at $m\tau=0$:
\begin{eqnarray}
\Lambda(m\tau) &\simeq & \frac{\log(m^2\tau^2)}{h} .
\end{eqnarray}

Similarly we obtain 
\begin{eqnarray}
\Lambda_\tau(m\tau) &\simeq& -\frac{6}{h}\left[1- \frac{3\sqrt{\pi}}{2}T^{1/2}+8T -\frac{25\sqrt{\pi}}{2}T^{3/2}\right], 
\label{eq:Lambda_tau3}\\
\frac{\delta \Lambda (m\tau)}{\Lambda_\tau(m\tau)}&\simeq&  
\frac{\sqrt{\pi}}{2}T^{1/2}+\frac{(3\pi-16)}{3}T 
+5\sqrt{\pi}\frac{(9\pi-20)}{24}T^{3/2}
\label{eq:del_Lambda_tau3}
\end{eqnarray}
at $d=3$ ,
\begin{eqnarray}
\Lambda_\tau(m\tau) &\simeq& -\frac{6}{h}\left[1+ T (2L+3)+T^2(9L+3) \right], 
\label{eq:Lambda_tau4}\\
\frac{\delta \Lambda (m\tau)}{\Lambda_\tau(m\tau)} &\simeq& 
-\frac{1}{2}\left\{T(2L+5)-T^2(10+7L+6L^2)\right\}
\label{eq:del_Lambda_tau4}
\end{eqnarray}
at $d=4$, and 
\begin{eqnarray}
\lim_{m\tau\rightarrow 0} \Lambda_\tau (m\tau) &=& -\frac{d(d-1)}{h(d-2)}, \qquad
\lim_{m\tau\rightarrow 0}\frac{\delta \Lambda (m\tau)}{\Lambda_\tau(m\tau)} = 0 
\end{eqnarray}
for general $d\ge 3$.
Therefore we have
\begin{eqnarray}
G_{\mu\nu}(\tau) &\simeq & -\Lambda g_{\mu\nu}(\tau), \qquad \Lambda \equiv-\frac{d(d-1)}{h(d-2)}, \qquad m\tau \ll 1,
\end{eqnarray}
which shows that the space becomes Euclidean AdS in the UV limit 
($m\tau \ll 1$) at $d\ge 3$.
These results are also consistent with results in the previous section.

\subsection{Behavior of the Einstein tensor from UV to IR}
If we define the radius of the AdS space $R$ as $R^2=-1/\Lambda $, we have
\begin{eqnarray}
R_{\rm UV}^2 =\frac{d-2}{d} R_{\rm IR}^2 < R_{\rm IR}^2
\end{eqnarray}
at $d \ge 3$. The radius increases toward the infrared.

\begin{figure}[tbh]
\begin{center}
\scalebox{0.5}{\includegraphics{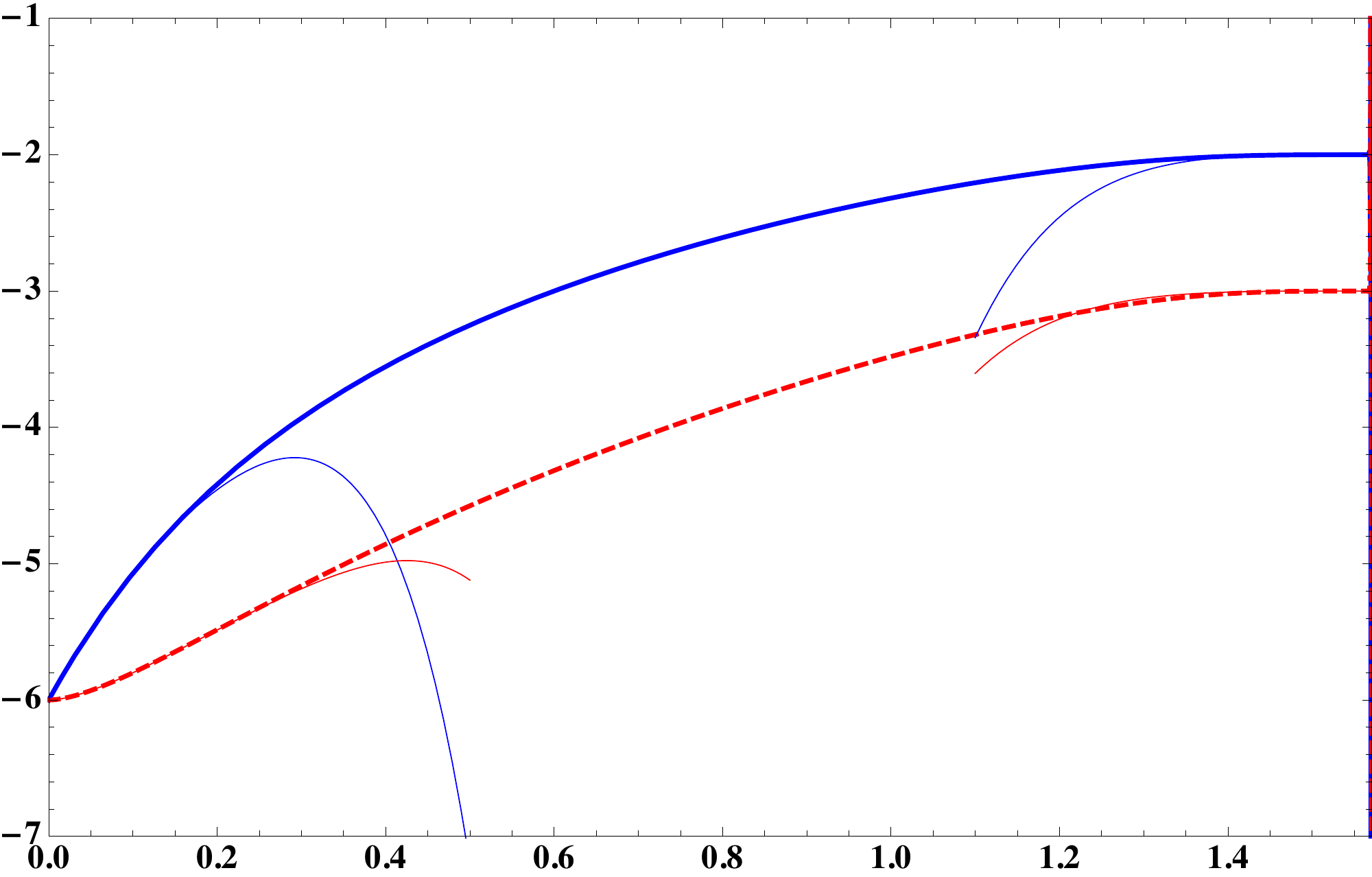}}
\end{center}
\caption{$h \Lambda_\tau(m\tau)$ as a function of $m\tau$ at $d=3$ (blue solid line) and $d=4$ (red dashed line), where the horizontal axis is $ x= \arctan (m\tau)$, together with asymptotic behaviors at UV ($x\simeq 0$) and IR ($x\simeq \pi/2$).
}
\label{fig:lambda}
\end{figure}

We show this behavior more explicitly at $d=3$ and 4, using the formula 
given in eqs.~(\ref{eq:Lambda_d}) and (\ref{eq:delta_Lambda}) .
In Fig.~\ref{fig:lambda}, $h\Lambda_\tau(m\tau)$ is plotted as a function 
of $m\tau$ at $d=3$ (blue solid line) and $d=4$ (red dashed line), 
where the horizontal axis is $ x= \arctan (m\tau)$,
together with asymptotic behaviors at UV ($x\simeq 0$) given in 
eqs.~(\ref{eq:Lambda_tau3}) and (\ref{eq:Lambda_tau4}), 
and those at IR ($x\simeq \pi/2$) given in eq.~(\ref{eq:Lambda_tau_IR}).
Starting from $h\Lambda_\tau(0) = -6$ in the UV limit, 
$h\Lambda_\tau(m\tau)$ increases toward the IR limit with 
$h\Lambda_\tau(\infty) = -2$ at $d=3$ and $h\Lambda_\tau(\infty) = -3$ at $d=4$.

\begin{figure}[tbh]
\begin{center}
\scalebox{0.5}{\includegraphics{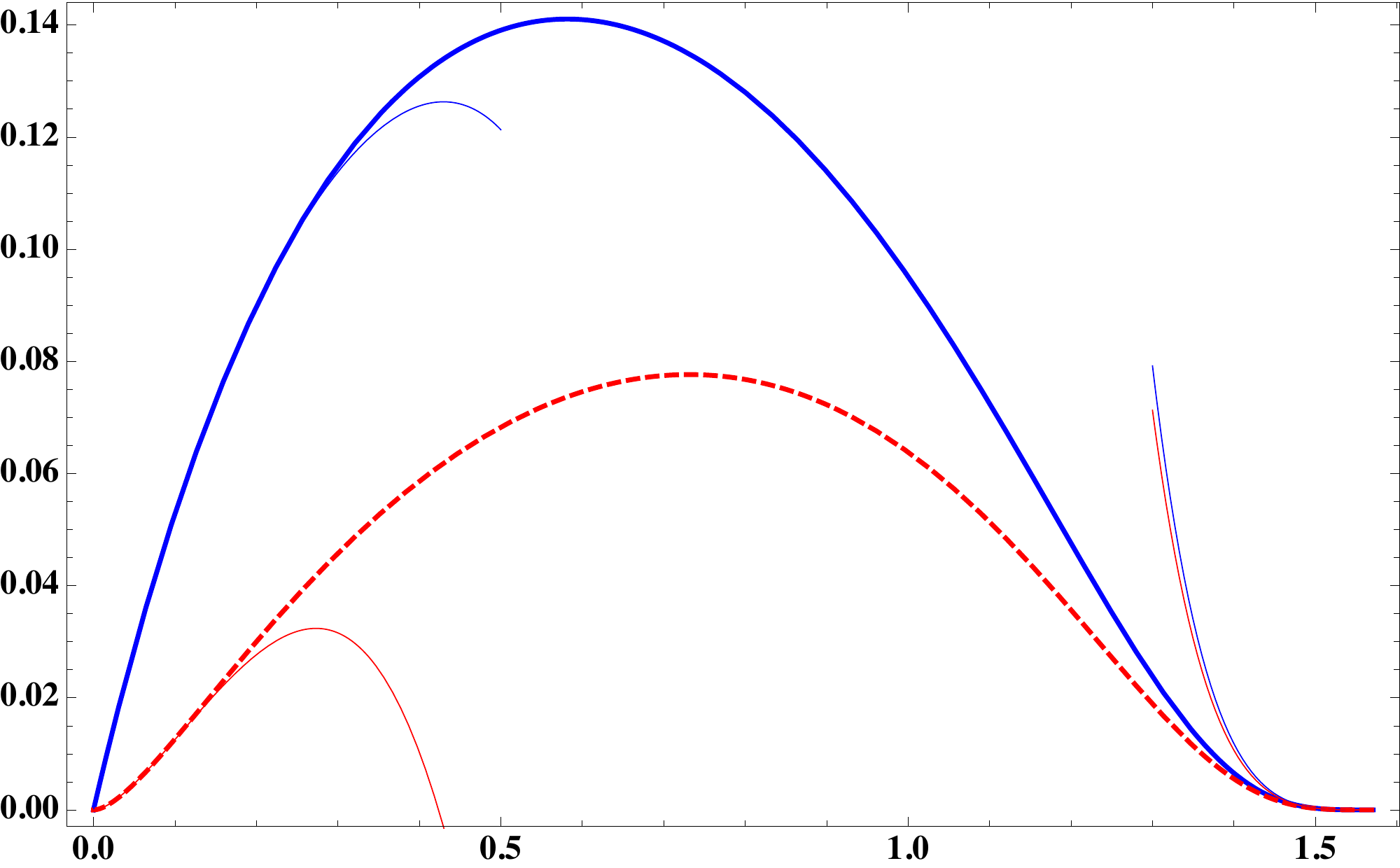}}
\end{center}
\caption{$\delta\Lambda(m\tau)/ \Lambda_\tau(m\tau)$ as a function of $m\tau$ at $d=3$ (blue solid line) and $d=4$ (red  dashed line), where the horizontal axis is $ x= \arctan (m\tau)$, together with asymptotic behaviors at UV ($x\simeq 0$) and IR ($x\simeq \pi/2$).
}
\label{fig:del_lambda}
\end{figure}

Fig.~\ref{fig:del_lambda}, on the other hand, shows that 
$\delta \Lambda(m\tau)/\Lambda_\tau(m\tau)$, which represents the violation 
of $\Lambda_\tau(m\tau) = \Lambda_d(m\tau)$, as a function of $m\tau$ 
at $d=3$ (blue solid line) and $d=4$ (red dashed line), where the horizontal 
axis is $ x= \arctan (m\tau)$,
together with asymptotic behaviors at UV ($x\simeq 0$) given in 
eqs.~(\ref{eq:del_Lambda_tau3}) and (\ref{eq:del_Lambda_tau4}), and those at 
IR ($x\simeq \pi/2$) given in eq.~(\ref{eq:del_Lambda_tau_IR}).
As expected, $\delta \Lambda(\tau)/\Lambda_\tau(m\tau)$ becomes zero 
in both UV and IR limits, and the violation becomes maximum 
at $m\tau \simeq 0.66$ ($x\simeq 0.58$) or at $m\tau \simeq 0.90$ ($x\simeq 0.73$), 
which is about 14\% at $d=3$ or 8\% at $d=4$, respectively.

\section{Summary}
\label{summary}
In this paper, we apply the method in Ref.~\cite{Aoki:2015dla} to the O($N$) 
invariant $\varphi^4$ model, where 
the $d+1$ dimensional metric is defined from the $d$ dimensional field theory 
through gradient flow in the large $N$ limit.
As generalizations of the proposal of Ref.~\cite{Aoki:2015dla}, 
we consider the case where the action for the flow equation is different 
from the action of the original theory. In addition, we have introduced 
the NLSM normalization for the flow field, with which the normalized flow 
field only depends on the renormalized mass of the original $d$ 
dimensional theory.
Using this normalized flow field, we define the $d+1$ dimensional 
induced metric, which is shown to describe a Euclidean AdS space 
in both UV and IR limits at $d\ge 3$.

The induced metric, and thus the geometry, from the flow field with 
the NLSM normalization, depends only on the renormalized mass $m$ in 
the original $d$ dimensional $\varphi^4$ theory, but neither bare parameters ($\mu^2, u$) separately nor flow parameters ($\mu_f^2,u_f$) at all. 
This uniqueness of the induced geometry from the NLSM flow may be natural, 
since the O($N$) $\varphi^4$ theory becomes a free massive field theory 
in the large $N$ limit, 
which does not depend explicitly on the bare coupling constant $u$ including 
the free and the NLSM limits.
In this sense, the large $N$ scalar field theory in $d$ dimensions 
corresponds to a $d+1$ dimensional classical geometry in the large $N$ 
limit, which becomes the Euclidean AdS in UV and IR limits at $d> 2$.
A posteriori, the NLSM normalization turns out to be an interesting 
choice to define the RG transformation, as an AdS space emerges 
in both UV and IR limits.

Since the information about the renormalized coupling constant appears 
at the NLO in the large $N$ expansion of the $d$ dimensional 
$\varphi^4$ model, our next task is to obtain the solution of the 
flow equation at NLO. Using the solution, we  then evaluate quantum 
corrections to the classical geometry such as the propagation of 
the induced metric $\hat g_{\mu\nu}$ on the classical background.
It is interesting to see how the information about interactions in 
the large $N$ field theory appear in quantum corrections to the induced metric.
Although it is rather difficult to solve the flow equation at 
NLO\cite{Aoki:2014dxa}, we are currently working on this problem.

\section*{Acknowledgement}
S. A thanks Dr. H. Suzuki for useful discussions.
He is supported in part by the Grant-in-Aid of the Japanese Ministry of Education, Sciences and Technology, Sports and Culture (MEXT) for Scientific Research (No. 25287046), by MEXT Strategic Program for Innovative Research (SPIRE) Field 5,  
by a priority issue (Elucidation of the fundamental laws and evolution of the universe) to be tackled by using Post ``K" Computer, 
and by Joint Institute for Computational Fundamental Science (JICFuS).
This investigation has also been supported in part by the Hungarian 
National Science Fund OTKA (under K116505).
S. A and J. B. would like to thank the Max-Planck-Institut f\"ur Physik 
for its kind hospitality during their stay for this research project.
T.O. is supported in part by the Grant-in-Aid of the Japanese 
Ministry of Education, Sciences and Technology, 
Sports and Culture (MEXT) for Scientific Research (No. 26400248).


%

\appendix

\section{Divergence of the flow field in perturbation theory}
\label{app_divergence}
In this appendix, we show that an extra divergence  appears in the flow field of the $\lambda \varphi^4$ theory, 
if the bare action is employed for the flow equation, together with perturbation theory in the coupling\cite{Capponi:2015ucc,Suzuki:2015}.
Although this kind of divergence can be avoided by using the renormalized flow equation
such as the free flow equation  in  \cite{Capponi:2015ucc,Monahan:2015lha}, 
it is relevant to our flow action, which contains bare parameters of the original theory.
We then discuss how this divergence disappears non-perturbatively in the large $N$ limit. 

\subsection{Renormalization in the original theory }
For simplicity, we consider the model at $d=2,3$ defined by
\begin{eqnarray}
S &=&\int \mrd^dx\,\left[ \frac{1}{2}\partial^k\varphi(x)\cdot\partial_k\varphi(x) +\frac{\mu^2}{2}\varphi^2(x) +\frac{\lambda}{4!}\varphi^2(x)^2\right],
\end{eqnarray}
where the renormalization is made by
\begin{eqnarray}
\mu^2 &=& Z_m m^2 = m^2 + \delta m^2, \quad
\varphi = Z_\varphi \varphi_R , \quad
\lambda = Z_\lambda \lambda_R .
\end{eqnarray}
At $d=2,3$, we have
\begin{eqnarray}
\delta m^2 &=& -\frac{N+2}{6}\lambda_R Z(m)+O(\lambda_R^2), \quad
Z_\varphi = 1+O(\lambda_R^2), \quad
Z_\lambda = 1+ O(\lambda_R^2).
\end{eqnarray}
Hereafter we neglect $O(\lambda_R^2)$ contributions. 
We then have
\begin{eqnarray}
\langle \varphi^a(x)\varphi^b(y)\rangle &=& \delta^{ab}\int \mrd p\, 
\frac{\mre^{ip(x-y)}}{p^2+m^2}.
\end{eqnarray}

\subsection{Divergence in the flowed field}
We now consider the flow equation, given by
\begin{eqnarray}
\dot\phi^a(t,x) &=&  (\Box -\mu^2) \phi^a(t,x) -\frac{\lambda}{6} \phi^a(t,x) \phi^2(t,x) ,  \quad \phi^a(0,x)=\varphi^a(x)
\end{eqnarray}
at $O(\lambda)$, where we take $\mu_f=\mu$ and $\lambda_f=\lambda$ for simplicity.

Setting $\phi =\phi_0 + \lambda \phi_1$, we obtain
\begin{eqnarray}
\phi_0^a(t,x) &=& \int \mrd y\, K_t(x-y)\varphi^a(y), \\
K_t(x) &=&\int \mrd p\, \mre^{ip x}\mre^{-t(p^2+m^2)}, 
\quad K_0(x) =\delta^{(d)}(x)\,,
\end{eqnarray}
and
\begin{eqnarray}
\lambda\phi_1^a(t,x) &=-& \int_0^t \mrd s\, \int \mrd y\, K_{t-s}(x-y) 
\phi_0^a(s,y) \left[ \delta m^2 +\frac{\lambda}{6}\phi_0^2(s,y)\right]\,.
\end{eqnarray}
 
The two-point function for the flow field is given by
\begin{eqnarray}
\langle \phi^a(t,x)\phi^b(s,y)\rangle &=& \langle \phi_0^a(t,x)\phi_0^b(s,y)\rangle \nonumber \\
&+&
 \langle \lambda\phi_1^a(t,x)\phi_0^b(s,y)\rangle +  \langle \phi_0^a(t,x) \lambda\phi_1^b(s,y)\rangle +O(\lambda^2),
\end{eqnarray}  
where we have
\begin{eqnarray}
 \langle \phi_0^a(t,x)\phi_0^b(s,y)\rangle &=& \delta^{ab}\int \mrd p\, 
\frac{\mre^{-(s+t)(p^2+m^2)+ip(x-y)}}{p^2+m^2}, \\
 \langle \lambda\phi_1^a(t,x)\phi_0^b(s,y)\rangle &=& -\delta m^2
\int_0^{t}\mrd t_1\,\int \mrd x_1\, K_{t-t_1}(x-x_1)  \langle \phi_0^a(t_1,x_1)\phi_0^b(s,y)\rangle  \nonumber \\
 &-&\frac{\lambda}{6}\int_0^{t}\mrd t_1\,\int \mrd x_1\, K_{t-t_1}(x-x_1)  
\langle \phi_0^a(t_1,x_1)\phi_0^2(t_1,x_1)\phi_0^b(s,y)\rangle .~~~~~~
\end{eqnarray}
The first term is evaluated as
\begin{eqnarray}
&=& -\delta^{ab}\delta m^2 \int_0^{t}\mrd t_1\,\int \mrd k\, 
\mre^{-(t-t_1)(k^2+m^2)+ik(x-x_1)}\int \mrd p\,
\frac{\mre^{-(t_1+s)(p^2+m^2)+ip(x_1-y)}}{p^2+m^2} \nonumber \\
&=& -\delta m^2 t  \langle \phi_0^a(t,x)\phi_0^b(s,y)\rangle,
\end{eqnarray}
while the second one is
\begin{eqnarray}
&=& -\frac{(N+2)\lambda}{6}\int_0^t \mrd t_1\, \int \mrd p\, 
\frac{\mre^{-2t_1(p^2+m^2)}}{p^2+m^2}  \langle \phi_0^a(t,x)\phi_0^b(s,y)\rangle
\nonumber \\
&=& \frac{(N+2)\lambda}{12} J_1(t)  \langle \phi_0^a(t,x)\phi_0^b(s,y)\rangle
\end{eqnarray}
where
\begin{eqnarray}
J_1(t) &\equiv & \int \mrd p\, \frac{\mre^{-2t(p^2+m^2)}-1}{(p^2+m^2)^2}
\end{eqnarray}
is UV finite at $d=2,3$.

We finally obtain
\begin{eqnarray}
\langle \phi^a(t,x)\phi^b(s,y)\rangle &=& \langle \phi_0^a(t,x)\phi_0^b(s,y)\rangle\nonumber\\
&\times&
\left[ 1 -(t+s)\delta m^2 +\frac{(N+2) \lambda}{12}\{J_1(t)+J_1(s)\}\right],
\label{eq:pert}
\end{eqnarray}
which shows that $t,s$ dependent renormalization is needed to 
make the two-point function for the flow field UV finite.
 
\subsection{Relation to the large $N$ result}
We now consider this divergence from the result in the large $N$ limit:
\begin{eqnarray}
\langle \phi^a(t,x)\phi^b(s,y)\rangle &=& \frac{\delta^{ab}}{N \sqrt{X(t)X(s)}} \int \mrd p\, \frac{\mre^{-(t+s) p^2+ip(x-y)}}{p^2+m^2},
\end{eqnarray} 
where 
\begin{eqnarray}
X(t) &=& \mre^{2t\mu^2}\left(1 + \frac{u}{3}\int_0^{t}\mrd x\,\int \mrd p\, 
\frac{\mre^{-2(p^2+\mu^2)x}}{p^2+m^2}\right) \nonumber \\
&=& \mre^{2t m^2}\left[ 1+ 2t \delta m^2 - \frac{u}{6} \int \mrd p\, 
\frac{\mre^{-2t (p^2+m^2)}-1}{(p^2+m^2)^2}\right] + O(\lambda^2).
\end{eqnarray} 
Therefore, if the perturbative expansion is employed, we have
\begin{eqnarray}
\langle \phi^a(t,x)\phi^b(s,y)\rangle &\simeq& \frac{\delta^{ab}
\mre^{-(t+s) m^2}}{N(1+(t+s) \delta m^2 - u \{J_1(t)+J_1(s)\}/12)} 
\int \mrd p\, \frac{\mre^{-(t+s) p^2+ip(x-y)}}{p^2+m^2} \nonumber \\
&\simeq&\frac{\delta^{ab}}{N}\left[1-(t+s) \delta m^2 +\frac{u}{12}\{J_1(t)+J_1(t)\} \right] \int \mrd p\, \frac{\mre^{-(t+s) (p^2+m^2)+ip(x-y)}}{p^2+m^2},\nonumber \\
\end{eqnarray}
which agrees with the result in eq.~(\ref{eq:pert}) in the large $N$ limit, 
after replacing $\lambda = u/N$ and $\phi \rightarrow \phi/\sqrt{N}$ 
in eq.~(\ref{eq:pert}).
This shows that the divergence which appeared in the perturbation 
expansion disappears in the large $N$ expansion where potentially 
divergent contributions are summed up to an exponential form.

\section{Incomplete gamma function}
\label{app_igamma}
The incomplete gamma function of the 2nd kind, $\Gamma(a, x)$, is defined by
\begin{eqnarray}
\Gamma(a, x) &=& \int_x^\infty \mrd y\, \mre^{-y} y^{a-1},
 \end{eqnarray}
whose asymptotic behaviors are given by
\begin{eqnarray}
\Gamma(a,x)&=& \Gamma(a)\mre^{-x}\left[\mre^{x} -  x^a\sum_{k=0}^\infty\frac{x^k}{\Gamma(k+1+a)}\right],
\quad x\rightarrow 0, \\
\Gamma(a,x) &=& x^{a-1}\mre^{-x}\sum_{k=0}^\infty \frac{\Gamma(a)}{\Gamma(a-k)}x^{-k}, \quad x\rightarrow\infty .
\end{eqnarray}
 
Other useful properties of $\Gamma$ are
\begin{eqnarray}
\Gamma(a+1,x) &=& a\Gamma(a,x) + x^a \mre^{-x},\quad  \Gamma(0,x) = - {\rm Ei}(-x) , \ x>0, \\
 \Gamma(1/2,x) &=& \sqrt{\pi}{\rm erfc}(\sqrt{x}), \quad \Gamma(1,x)=\mre^{-x},
\end{eqnarray}
where ${\rm erfc}$ is a complementary error function defined by
\begin{eqnarray}
{\rm erfc}(x) &=& \frac{2}{\sqrt{\pi}}\int_x^\infty \mrd y\, \mre^{-y^2} , \qquad {\rm erfc}(0) = 1,
\end{eqnarray}
and ${\rm Ei}(-x)$ is the exponential integral function defined by
\begin{eqnarray}
{\rm Ei}(-x) = -\int_x^\infty \mrd y\, \frac{\mre^{-y}}{y} .
\end{eqnarray}
As $x\rightarrow 0^+$, we have
\begin{eqnarray}
{\rm Ei}(-x) &=& \gamma + \log(x) +\sum_{n=1}^\infty \frac{(-x)^n}{n\, n!}, \quad
{\rm erfc}(x) = 1 -\frac{2}{\sqrt{\pi}} \sum_{n=0}^\infty \frac{(-1)^n x^{2n+1}}{(2n+1)\, n!},~~~~~
\end{eqnarray}
while as $x\rightarrow \infty$
\begin{eqnarray}
{\rm Ei}(-x) &=&\mre^{-x}\sum_{n=1}^\infty \frac{(n-1)!}{(-x)^{n}}, \quad
{\rm erfc}(x) = \frac{\mre^{-x^2}}{\sqrt{\pi} x}\left(1+ \sum_{n=1}^\infty \frac{(-1)^n (2n-1)!!}{(2x^2)^n}\right).~~~~
\end{eqnarray}

Using the above formulae, we obtain the asymptotic behavior of 
$\Gamma(1-d/2, 2m^2t)$. As $t\rightarrow \infty$, we have
\begin{eqnarray}
\Gamma(1-\frac d2, 2m^2t) &=& \frac{1}{(2t)^{d/2} m^d}\mre^{-2m^2 t}\sum_{k=0}^\infty\frac{\Gamma(1-\frac{d}{2})}{\Gamma(1-\frac{d}{2}-k)}\frac{1}{(2m^2t)^k},
\end{eqnarray}
while as $t\rightarrow 0$, 
\begin{eqnarray}
\Gamma(1-\frac d2, 2m^2t) &=&\left\{
\begin{array}{lc}
\displaystyle \sqrt{\pi} -2\sum_{k=0}^\infty \frac{(-1)^k (2m^2t)^{k+1/2}}
{(2k+1) k!}\,,  &   d=1\,,   \\
\\
\displaystyle  -\gamma - \log(2m^2t) -\sum_{k=1}^\infty
\frac{ (-2m^2t)^k}{k k!}\,,  &   d=2\,,   \\
\\
\displaystyle \frac{2\mre^{-2m^2t}}{(2m^2t)^{1/2}} 
-2\sqrt{\pi}{\rm erfc}(\sqrt{2m^2t})\,, &   d=3\,, \\
\\
\displaystyle  \frac{\mre^{-2m^2t}}{2m^2t} +{\rm Ei}(-2m^2t)\,, &  d=4\,.
\end{array}
\right. 
\end{eqnarray}

\end{document}